\documentclass{llncs}
\usepackage{graphicx}

\usepackage{stmaryrd}
\usepackage{mdframed}
\usepackage{mathpartir}
\usepackage{amsmath}
\usepackage{amssymb,makecell}
\usepackage[bookmarks=true,colorlinks=true, linkcolor=red, citecolor=cyan]{hyperref}
\usepackage[color,all,poly,matrix,arrow]{xy}

\newcommand{\LTO}[1]{\overset{\sf{#1}}{\bullet}}

\newcommand{\LTOO}[1]{\stackrel{\sf{#1}}{\circ}}
\newcommand{\DTOO}[1]{\underset{\sf{#1}}{\circ}}

\newcommand{\Too}[1]{\xrightarrow{\ \ #1\ \ }}

\newcommand{\Fromm}[1]{\xleftarrow{\ \ #1\ \ }}

\begin{document}

\def\Inst{{\bf Inst}}
\def\iinst{{{ \textendash}\bf \Inst}}
\def\taking{:}

\title{Algebraic Model Management: A Survey}
\author{
	Patrick Schultz\inst{1} \and David I. Spivak\inst{1} \and
    Ryan Wisnesky\inst{2}
}
\institute{Massachusetts Institute of Technology \and Conexus AI}
\maketitle              

\begin{abstract}
We survey the field of model management and describe a new model management approach based on algebraic specification.
\end{abstract}

%
\section{Introduction}

In this paper we survey the field of {\it model management} and describe a new model management approach based on techniques from the field of {\it algebraic specification}, with the hope of establishing an interlingua between the two fields.  By ``model management'' we mean ``meta-data intensive'' database management in the sense of Bernstein \& Melnik~\cite{Bernstein:2007:MMM:1247480.1247482}, which we define in Section~\ref{mm}.  By ``a new algebraic model management approach'' we mean our particular way~\cite{patrick}~\cite{jfp} of specifying database schemas and instances using algebraic (equational) theories.  

We first noticed a connection between model management and algebraic specification while investigating applications of {\it category theory}~\cite{BW} to {\it data integration}~\cite{Doan:2012:PDI:2401764}.  These investigations are described in~\cite{patrick} and~\cite{jfp}, and we present no substantial new results in this paper.  We assume readers have basic proficiency with category theory~\cite{BW}, algebraic specification~\cite{Mitchell:1996a}, and SQL.

{\bf Outline.} In Section~\ref{mm} we describe the traditional approach to model management and in Section~\ref{main} we describe our algebraic approach.  Also in Section~\ref{main} we describe the open-source CQL (Categorical Query Language) tool, available for download at \url{http://categoricaldata.net}, which implements our approach in software.  We conclude in Section~\ref{concl} by comparing our approach with the traditional approach. 

\section{Model Management}
\label{mm}
To quote from Melnik~\cite{Melnik:2004:GMM:1007755}:
\begin{quote}
Many challenging problems facing information systems engineering involve the manipulation of complex metadata artifacts, or \emph{models}, such as database schemas, interface specifications, or object diagrams, and mappings between models. The applications that solve metadata manipulation problems are complex and hard to build. The goal of generic model management is to reduce the amount of programming needed to develop such applications by providing a database infrastructure in which a set of high-level algebraic operators, such as Match, Merge, and Compose, are applied to models and mappings as a whole rather than to their individual building blocks.
 \end{quote}
 
In the paragraph above the word ``model'' is defined to mean a metadata artifact such as a schema, which conflicts with the definition of the word ``model'' as a structure satisfying a theory.  In this paper, we use the phrase ``model management'' to mean the field identified above, and use the word ``model'' to mean a structure satisfying a theory. 

Today model management is a large sub-field of information management with a research literature containing hundreds of published articles~\cite{Bernstein:2007:MMM:1247480.1247482}.  There is a consensus in that literature~\cite{Bernstein:2007:MMM:1247480.1247482} that model management is concerned with at least the  problems described in the next sections.

\subsection{Schema mapping} 
\label{secmap1}
Given two database schemas $S$ and $T$, the {\it schema mapping} problem~\cite{FKMP05} is to construct a ``mapping'' $F : S \to T$ that captures some user-specified relationship between $S$ and $T$.  Different model management systems use different notions of schema, including SQL, XML, and RDF~\cite{Bernstein:2007:MMM:1247480.1247482}.  The most common mapping formalism studied in the literature is that of ``embedded dependencies'' (EDs)~\cite{Doan:2012:PDI:2401764}: formulae in a fragment of first-order logic with useful computational properties.

We will use SQL schemas and EDs in our examples in this section.  Consider the following SQL schema $S$, consisting of two tables connected by a foreign key:
\begin{verbatim}
CREATE TABLE N2(ID INT PRIMARY KEY, age INT)

CREATE TABLE N1(ID INT PRIMARY KEY, name STRING, salary INT, 
     f INT FOREIGN KEY REFERENCES N2(ID)) 
\end{verbatim}
and the following SQL schema $T$, consisting of one table:
\begin{verbatim}
CREATE TABLE N(ID INT PRIMARY KEY, age INT, 
     name STRING, salary INT). 
\end{verbatim}
These two SQL schemas are displayed graphically in Figure~\ref{fkmex}.  

\begin{figure}[h]
\begin{mdframed}
\vspace{-.1in}
\begin{centering}
\begin{align*} 
\parbox{1.2in}{\fbox{\xymatrix@=8pt{
& \LTOO{String}\\
\LTO{N1} \ar[ur]^{\sf name} \ar[dr]_{\sf salary} \ar[rr]^{\sf f} & & \LTO{N2}\ar[dl]^{\sf age}\\
&  \DTOO{Int}}} }
\Too{F}
\parbox{1.2in}{\fbox{\xymatrix@=8pt{
\LTOO{String}\\
\LTO{N}\ar[u]^{\sf name} \ar@/^/[d]^{\sf age} \ar@/_/[d]_{\sf salary}\\
\DTOO{Int}}}}
\end{align*}
\end{centering}


\begin{footnotesize}

\begin{tabular}{cccc}
\multicolumn{4}{c}{{\sf N1}} \vspace{.01in} \\
 {\sf ID}  &{\sf name}&  {\sf salary} &  {\sf f}   \\\hline 
1&Alice&\$100 & 1 \\\hline 
2&Bob&\$250 & 2 \\\hline 
3&Sue&\$300 & 3 \\
\end{tabular}
\hspace{.2in}
\begin{tabular}{cc}
\multicolumn{2}{c}{{\sf N2}} \vspace{.01in} \\
 {\sf ID}  &   {\sf age}   \\\hline 
1&20\\\hline 
2&20\\\hline 
3&30\\
\end{tabular}
\hspace{.12in}
$\overset{\Fromm{\llbracket \Delta_F \rrbracket }}{\Too{\llbracket \Pi_F \rrbracket, \llbracket \Sigma_F \rrbracket }}$
\hspace{.12in}
\begin{tabular}{cccc}
\multicolumn{4}{c}{{\sf N}} \vspace{.01in} \\
 {\sf ID}  &   {\sf name}   &   {\sf salary}   &   {\sf age}   \\\hline 
1&Alice&\$100&20\\\hline 
2&Bob&\$250&20\\\hline 
3&Sue&\$300&30\\
\end{tabular}
\end{footnotesize}


\caption{Example Data Migrations, with Foreign Keys (see Sections~\ref{secmap1},~\ref{schmap})}
\label{fkmex}
\end{mdframed}
\end{figure}

An example schema mapping $F : S \to T$ expressing that the target table {\sf N} is the join of source tables {\sf N1} and {\sf N2} along the column {\sf f} is:
$$
\forall id_1, id_2, a, n, s. \ {\sf N1}(id_1, n, s, id_2) \wedge {\sf N2}(id_2, a) \to {\sf N}(id_1, a, n, s). 
$$
Two instances satisfying the above ED are shown in Figure~\ref{fkmex}.  In general, many EDs can map between two SQL schemas. \\

\subsection{Query generation}  
\label{qgen}

Given a schema mapping $F : S \to T$, the {\it query generation} problem~\cite{Bernstein:2007:MMM:1247480.1247482} is to construct a query which converts databases on $S$ to databases on $T$ in a way that satisfies $F$.  The query languages typically studied include SQL, XQuery, and various comprehension- and $\lambda$-calculi~\cite{Bernstein:2007:MMM:1247480.1247482}.  

A SQL query to implement the example mapping from Section~\ref{secmap1} is:
\vspace{-.4in}
\begin{verbatim}
INSERT INTO N
SELECT N1.ID, N1.age, N2.name, N1.sal  
FROM N1, N2
WHERE N1.f = N2.ID
\end{verbatim}
Technically, the \texttt{INSERT} portion of the above SQL code is not a ``query'', but rather an ``update'', and in practice the code generated from a query generation task will often store the results of the query.  An example of running the above SQL is shown as the left-to-right direction of Figure~\ref{fkmex}.  In general, many or no SQL queries may implement a set of EDs~\cite{Doan:2012:PDI:2401764}.  EDs can also be directly executed by an algorithm called ``the chase''~\cite{Doan:2012:PDI:2401764}. 

\subsection{Mapping Inversion}
\label{secinv1}

Given a schema mapping $F : S \to T$, the {\it mapping inversion} problem~\cite{Fagin:2007:ISM:1292609.1292615} is to construct a schema mapping $F^{-1} : T \to S$ that undoes $F$ with respect to query generation (i.e. the queries generated from $F$ and $F^{-1}$ should be inverses). 

The natural candidate ED to invert the schema mapping of Section~\ref{secmap1} expresses that {\sf N} projects onto {\sf N1} and {\sf N2}:
$$
\forall id_1, a, n, s. \ {\sf N}(id_1, a, n, s) \to \exists id_2 . \ {\sf N1}(id, n, s, id_2) \wedge {\sf N2}(id_2, a)
$$
and a possible SQL implementation of this ED is: \\

\begin{minipage}[t]{.5\textwidth}
\begin{verbatim}
INSERT INTO N1
SELECT ID, name, sal, ID  
FROM N
\end{verbatim}
\end{minipage}
\vrule{}\hspace{.1in}
\begin{minipage}[t]{.5\textwidth}
\begin{verbatim}
INSERT INTO N2
SELECT ID, age  
FROM N
\end{verbatim}
\end{minipage}
$$ $$
However, the above ED is \emph{not} an inverse to the ED of Section~\ref{secmap1}, as is seen by taking $\emptyset = {\sf N1} \neq {\sf N2}$.   Indeed, it is rare for an ED, or set of EDs, to be invertible, and weaker notions of inverse, such as ``quasi-inverse''~\cite{Fagin:2007:ISM:1292609.1292615}, are common in the literature~\cite{Fagin:2007:ISM:1292609.1292615}.   An example of running the above SQL is shown as the right-to-left direction of Figure~\ref{fkmex}.  

\subsection{Mapping Composition}
\label{comp}

Given schema mappings $F: S \to T$ and $G: T \to U$, the {\it mapping composition} problem~\cite{Fagin:2005:CSM:1114244.1114249} is to construct a schema mapping $G \circ F : S \to U$ that is equivalent with respect to the query generation problem (i.e. running the query generated from $G \circ F$ should have the same effect as running the query generated from $G$ on the results of the query generated from $F$). 

The composition of the ED from Section~\ref{secmap1} with the ED from Section~\ref{secinv1} is
$$
\forall id_1, id_2, n, s, a. \ {\sf N1}(id_1, n, s, id_2) \wedge {\sf N2}(id_2, a) \to \exists x . \ {\sf N1'}(id, n, s, x) \wedge {\sf N2'}(x, a)
$$
where {\sf N1'}, {\sf N2'} are target ``copies'' of source tables {\sf N1}, {\sf N2}.  This composed ED is not the identity, thereby showing that the ED from Section~\ref{secinv1} does not invert the ED from Section~\ref{secmap1}.  In the case of EDs, composed mappings may not exist~\cite{Fagin:2005:CSM:1114244.1114249}, but some restrictions and extensions of EDs are closed under composition~\cite{Fagin:2005:CSM:1114244.1114249}.


\subsection{Schema matching}
\label{matching}
  Given two database schemas $S$ and $T$, the {\it schema matching} problem~\cite{Doan:2012:PDI:2401764} is to automatically find ``correspondences'' between $S$ and $T$ and to automatically infer schema mappings $S \to T$ from these correspondences.  In general, inference of entire mappings cannot be fully automated and the focus of the matching problem is to reduce the human effort required to construct a schema mapping by e.g., suggesting partial mappings that can be completed by users.  There are many techniques for schema matching ranging from comparison of column names by string similarity to machine learning algorithms; for an overview, see~\cite{Doan:2012:PDI:2401764}.  In the example from Section~\ref{secmap1}, two correspondences that are easy to automatically find are $({\sf N1}, {\sf N})$ and $({\sf N2}, {\sf N})$ and tools such as Clio~\cite{Haas:2005:CGU:1066157.1066252} can create the ED from Section~\ref{secmap1} from these two correspondences. \\


\subsection{Further References}

In this paper we will focus on the problems described in the previous sections, but many other problems are studied in the model management literature~\cite{Bernstein:2007:MMM:1247480.1247482}, and many of these problems are related to algebraic specification.  For example, \emph{schema/instance merge} problems~\cite{Bernstein:2007:MMM:1247480.1247482}, which arise often in data integration scenarios~\cite{DBLP:conf/dbpl/AlagicB01}, can be formalized as pushouts in suitable categories of schemas/instances~\cite{jfp}, and such pushouts are related to model-theoretic concepts such as {\it model amalgamation}~\cite{Hodges:1997:SMT:262326}.

\begin{figure}
\hspace{.05in}
\includegraphics[width=4.7in]{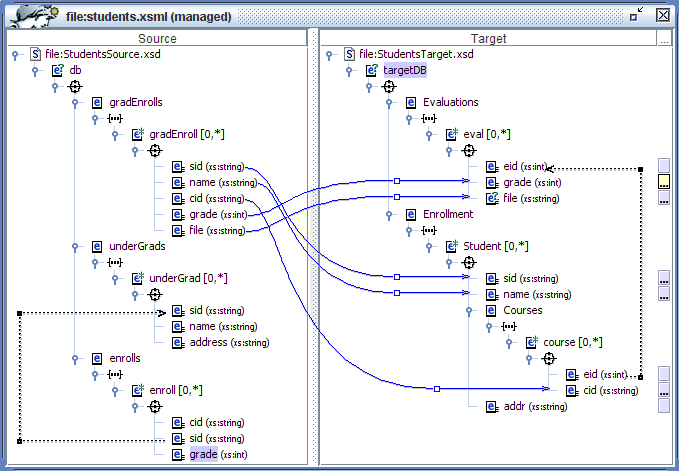}
\caption{A schema mapping in Clio~\cite{Haas:2005:CGU:1066157.1066252}}
\label{clio}
\end{figure}

Many software products solve model management problems~\cite{Bernstein:2007:MMM:1247480.1247482}, including {\it ETL (Extract, Transform, Load)} tools~\cite{Doan:2012:PDI:2401764}, which extract data from separate databases, apply user-specified transformations, and then load the result into a target system such as a data warehouse; {\it query mediators}~\cite{Doan:2012:PDI:2401764}, which answer queries about a ``virtual'' integrated database by combining queries about separate source databases; and {\it visual schema mapping tools}~\cite{Haas:2005:CGU:1066157.1066252} which allow users to create schema mappings by visually connecting related schema elements with lines, as shown in Figure~\ref{clio}.

There have been at least two attempts to provide a ``meta semantics'' for model management operations.   In~\cite{Melnik:2004:GMM:1007755} Melnik gives a ``state based''  meta semantics to some of the above operations by defining a schema mapping $S \to T$ to be an arbitrary binary relation between instances on $S$ and instances on $T$; the ED-based semantics described above is an instantiation of this meta semantics.  In~\cite{DBLP:conf/dbpl/AlagicB01} and~\cite{Goguen04informationintegration} the authors give an ``institution theoretic'' meta semantics to some of the above operations by defining a schema mapping $S \to T$ to be a morphism in a suitable category of schemas; CQL's semantics is an instantiation of this meta semantics.  


\section{Algebraic Model Management}
\label{main}

Our approach to model management is based on the algebraic approach to databases, data migration, and data integration we describe in~\cite{patrick} and~\cite{jfp}.  Those works, and hence this work, extend a particular category-theoretic data model that originated in the late 1990s~\cite{Fleming02adatabase} and was later extended in~\cite{Spivak:2012:FDM:2324905.2325108} and~\cite{relfound} and implemented in CQL  (\url{http://categoricaldata.net}).  

In the next section we describe our formalism for database schemas and instances and introduce CQL.  The subsequent sections implement the model management operations from Section~\ref{mm} using our formalism.  In this section we abbreviate ``algebraic theory'' as ``theory''.


\subsection{Algebraic Databases}
\label{algdb}

In our formalism~\cite{jfp}, database schemas and instances are defined as theories of a certain kind, which we describe in the next sections.  For ease of exposition, we will sometimes conflate schemas and instances as defined in our formalism with their CQL equivalents.

\vspace{-.1in}
\subsubsection{Type sides}
We first fix a theory, $\mathit{Ty}$, called the {\it type side} of our formalism. The sorts of $\mathit{Ty}$ are called {\it types} and the functions of $\mathit{Ty}$ are the functions that can appear in schemas and instances.  

CQL allows arbitrary theories to be used as type sides.  But we have found that in practice, CQL users almost always want to use the theory of an existing programming language, say java, for their type side.  The ability to ``bind'' CQL to an existing language is particularly important in model management because input data may only be accessible through, e.g., a java API.  For this reason, CQL allows a type side to be defined by specifying, for each sort $s$, a java class $C_s$ and a java function ${\sf String} \to C_s$ that tells CQL how to interpret the strings it encounters in CQL programs as objects of $C_s$. 

An example CQL type side about integers and strings is shown in Figure~\ref{CQLty}.  This type side defines a theory with two sorts and infinitely many constants -- all the java strings and integers -- and no equations.  The java code for \texttt{Int} says that whenever a string $x$ is encountered in an CQL program and a term of sort \texttt{Int} is required, that java's \texttt{parseInt} function should be applied to $x$ to yield the desired \texttt{Int}.  The keyword \texttt{literal}, used in many places in CQL, indicates a literal (user-defined constant) definition.

\begin{figure}
\begin{mdframed}
\begin{verbatim}
typeside Ty = literal { 
    java_types
         String = "java.lang.String"
         Int = "java.lang.Integer"
    java_constants
         String = "return input[0]"
         Int = "return java.lang.Integer.parseInt(input[0])"
}
\end{verbatim}
\vspace{-.1in}
\caption{CQL type side $\mathit{Ty}$}
\label{CQLty}
\end{mdframed}
\end{figure}

\vspace{-.1in}
\subsubsection{Schemas}

A {\it schema} on type side $\mathit{Ty}$ is a theory extending $\mathit{Ty}$ with new sorts (called {\it entities}), new unary functions from entities to types (called {\it attributes}), new unary functions from entities to entities (called {\it foreign keys}), and new equations (called {\it data integrity constraints}) of the form $\forall v:s. \ t = t'$, where $s$ is an entity and $t,t'$ are terms of the same type, each containing a single free variable $v$.  The restrictions in the preceding sentence (e.g., no functions from types to entities) are necessary to use our formalism for model management purposes~\cite{patrick}~\cite{jfp}.  

Figure~\ref{CQLsch} shows the CQL schemas corresponding to Figure~\ref{fkmex}.  These schemas contain no equations and are both on the type side $\mathit{Ty}$ defined in Figure~\ref{CQLty}.  

\begin{figure}
\begin{mdframed}
\begin{minipage}[t]{.5\textwidth}
\begin{verbatim}
schema S = literal : Ty {
    entities 
         N1 
         N2
    foreign_keys	
         f : N1 -> N2
    attributes
         name : N1 -> String
         salary : N1 -> Int
         age : N2 -> Int
}
\end{verbatim} 
\end{minipage}
\vrule{} \hspace{.1in}
\begin{minipage}[t]{.5\textwidth}
\begin{verbatim}
schema T = literal : Ty {
    entities 
         N
    attributes
         name : N -> String
         salary : N -> Int
         age : N -> Int
}
\end{verbatim}
\end{minipage}
\vspace{-.1in}
\caption{CQL schemas $S$ and $T$ on type side $\mathit{Ty}$}
\label{CQLsch}
\end{mdframed}
\end{figure}

\begin{figure}
\begin{mdframed}
\begin{verbatim}
instance I = literal : S {
    generators 
         1 2 3 : N1
    equations
         name(1) = Alice  salary(1) = 100   age(f(1)) = 20
         name(2) = Bob    salary(2) = 250   age(f(2)) = 20
         name(3) = Sue    salary(3) = 300   age(f(3)) = 30
}  
\end{verbatim}
\vspace{-.1in}
\caption{CQL instance $I$ on schema $S$}
\label{CQLinst}
\end{mdframed}
\end{figure}

\begin{figure}
\hspace{-.3in}
\includegraphics[width=5.35in]{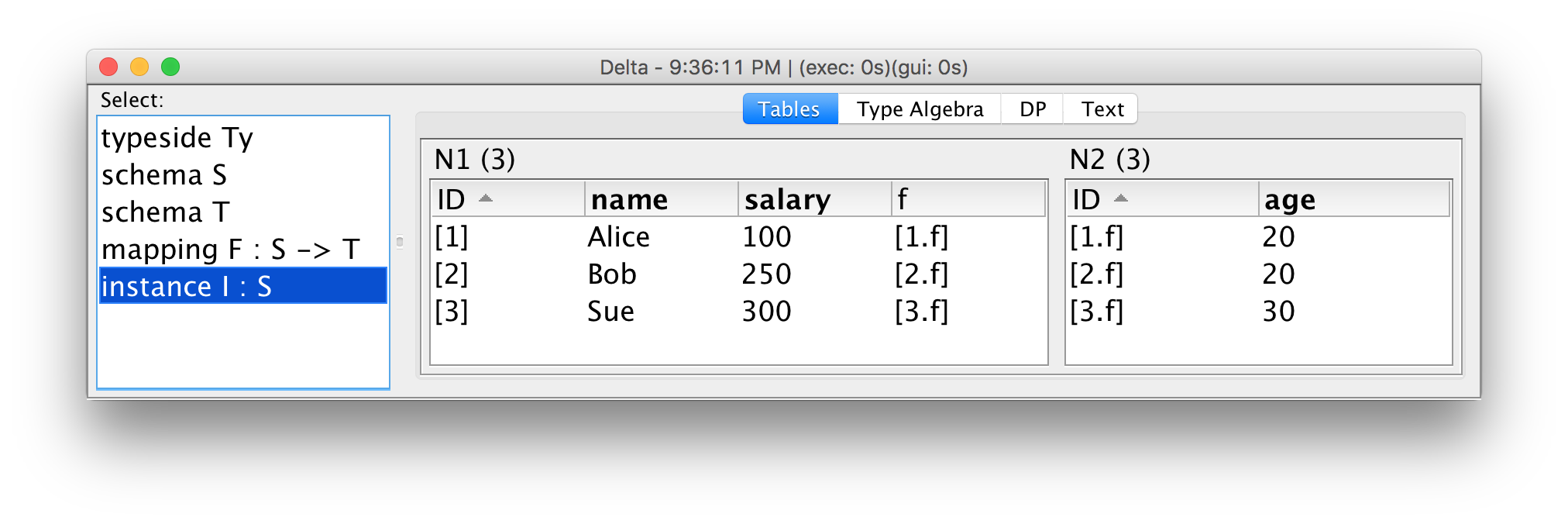}
\vspace{-.5in}
\caption{Initial algebra for CQL instance $I$}
\label{CQLfig}
\end{figure}

\subsubsection{Instances}
\label{secinst}

An {\it instance} $I$ on schema $S$ is a theory extending $S$ with new 0-ary function (constant) symbols called {\it generators} and non-quantified equations.  An example CQL instance on schema $S$ (Figure~\ref{CQLsch}) is shown in Figure~\ref{CQLinst}.  

The intended meaning of an instance $I$, written $\llbracket I \rrbracket$, is the {\it term model} (i.e., {\it initial algebra}) for $I$ which contains, for each sort $s$, a {\it carrier set} consisting of the closed terms of sort $s$ modulo provability in $I$.  A morphism of instances $I \to J$ is  a homomorphism (natural transformation) of algebras $\llbracket I \rrbracket \to \llbracket J \rrbracket$.

 Figure~\ref{CQLfig} shows the meaning of the instance $I$ from Figure~\ref{CQLinst} in the CQL tool.   The CQL tool visually displays term models as sets of tables, one per entity $e$, each with an ID column corresponding to the carrier set for $e$.  The tables in Figure~\ref{CQLfig} are isomorphic to the left tables in Figure~\ref{fkmex}.
 

In the following sections we implement the model management operations from Section~\ref{mm} using the preceding definitions of schema and instance. 


\vspace{-.1in}
\subsection{Schema mapping}  
\label{schmap}

Given schemas $S,T$, the {\it schema mapping} problem (Section~\ref{secmap1}) is to construct a ``mapping'' $F : S \to T$ that captures some relationship between $S$ and $T$.  

Let $S$ and $T$ be CQL schemas on the same type side $\mathit{Ty}$.  An CQL schema mapping $F : S \to T$ is defined as a ``derived signature morphism''~\cite{Mossakowski2015} from $S$ to $T$ that is the identity on $\mathit{Ty}$.  That is, $F : S \to T$ assigns to each entity $e \in S$ an entity $F(e) \in T$, and to each attribute / foreign key $f : s \to s'$ a term $F(f)$, of type $F(s')$ and with one free variable of type $F(s)$, in a way that respects  equality: if $S \vdash t = t'$, then $T \vdash F(t) = F(t')$.  We have found that many mappings arising in practice cannot be expressed using plain signature morphisms and require the more general notion of ``derived'' signature morphism.  

Whereas a schema mapping in Section~\ref{secmap1} was an ED (formula in a fragment of first-order logic), which induces a single binary satisfaction relation between instances, CQL schema mappings are derived signature morphisms and induce three relations between instances, which we will describe in the next section.

An example CQL schema mapping $F : S \to T$ is shown in Figure~\ref{CQLmapX}, where CQL schemas $S$ and $T$ are defined in Figure~\ref{CQLsch}.  This mapping is also shown graphically in Figure~\ref{fkmex}. 

\begin{figure}[h]
\begin{mdframed}
\begin{verbatim}
mapping F = literal : S -> T {
    entities 
         N1 -> N
         N2 -> N
    foreign_keys
         f -> lambda x:N. x 
    attributes
         name -> lambda x:N. name(x)
         salary -> lambda x:N. salary(x)
         age -> lambda x:N. age(x)
} 
\end{verbatim}\vspace{-.1in}
\caption{CQL schema mapping $F : S \to T$}
\label{CQLmapX}
\end{mdframed}
\vspace{-.2in}
\end{figure}



%

\vspace{-.1in}
\subsection{Query generation}  

Given a mapping $F : S \to T$, the {\it query generation} problem (Section~\ref{qgen}) is to use $F$ to construct a query which converts databases on $S$ to databases on $T$.

In our formalism, the database instances and morphisms on a schema $S$ constitute a category, denoted $S\iinst$, and a schema mapping $F : S \to T$ induces a functor $\Sigma_F : S\iinst \to T\iinst$ defined by substitution.  The functor $\Sigma_F$ has a right adjoint, $\Delta_F : T\iinst \to S\iinst$, which corresponds to the ``model reduct functor'' when our formalism is described in institution-theoretic terms~\cite{DBLP:conf/dbpl/AlagicB01}. The functor $\Delta_F$ has a right adjoint, $\Pi_F : S\iinst \to T\iinst$.  See~\cite{patrick} for proof that $\Delta_F$ always has left and right adjoints.   As adjoints, $\Delta_F,\Pi_F$ preserve limits and $\Delta_F,\Sigma_F$ preserve colimits, implying many useful properties; for example, $\Sigma_F(I + J) \cong \Sigma_F(I) + \Sigma_F(J)$ and $\Pi_F(I \times J) \cong \Pi_F(I) \times \Pi_F(J)$.  

Note that unlike Section~\ref{secmap1}, where there was a single query associated with a schema mapping (ED), in our algebraic approach there are three queries, one for each of $\Delta_F$, $\Sigma_F$, $\Pi_F$.   The conditions under which $\Delta_F$,$\Sigma_F$,  $\Pi_F$ can be expressed in SQL and vice-versa are characterized in~\cite{relfound}.  

Although it is possible to give explicit formulae to define $\Delta_F, \Sigma_F, \Pi_F$~\cite{patrick} we instead give examples in Figures~\ref{fkmex} and~\ref{fmd}.  Note that in these examples we are not showing instances (theories) as defined in Section~\ref{algdb}; we are showing term models.   For this reason, we surround $\Delta_F,\Sigma_F,\Pi_F$ with denotation brackets $\llbracket \rrbracket$ in these examples.  In addition, as adjoints $\Delta,\Sigma,\Pi$ are only defined up to unique isomorphism, so we arbitrarily make up names for IDs and in these examples.  
Figures~\ref{fkmex} and~\ref{fmd} show an CQL schema mapping $F$ which takes two distinct source entities, {\sf N1} and {\sf N2}, to the target entity {\sf N}.  The $\llbracket \Delta_F \rrbracket$ functor projects in the opposite direction of $F$: it projects columns from the single table for {\sf N} to two separate tables for {\sf N1} and {\sf N2}, similar to {\tt FROM N AS N1} and {\tt FROM N AS N2} in SQL.  When there is a foreign key from {\sf N1} to {\sf N2}, the $\llbracket \Delta_F \rrbracket$ functor populates it so that {\sf N} can be recovered by joining {\sf N1} and {\sf N2}.  The $\llbracket \Pi_F \rrbracket$ functor takes the cartesian product of {\sf N1} and {\sf N2} when there is no foreign key between {\sf N1} and {\sf N2}, and joins {\sf N1} and {\sf N2} along the foreign key when there is.  The $\llbracket \Sigma_F \rrbracket$ functor disjointly unions {\sf N1} and {\sf N2}; because {\sf N1} and {\sf N2} are not union compatible (have different columns), $\llbracket \Sigma_F \rrbracket$ creates null values.  When there is a foreign key between {\sf N1} and {\sf N2}, $\llbracket \Sigma_F \rrbracket$ merges the tuples that are related by the foreign key, resulting in a join.  As these examples illustrate, $\Delta_F$ can be thought of as projection, $\Pi_F$ can be thought of as a product followed by a filter (which can result in a join), and $\Sigma_F$ can be thought of as a disjoint union (which does not require union-compatibility) followed by a merge (which can also result in a join).   

\begin{figure}[hp]

\begin{centering}
\begin{align*} 
\parbox{1.2in}{\fbox{\xymatrix@=8pt{
& \LTOO{String}\\
\LTO{N1} \ar[ur]^{\sf name} \ar[dr]_{\sf salary} & & \LTO{N2}\ar[dl]^{\sf age}\\
&  \DTOO{Int}}} }
\Too{F}
\parbox{1.2in}{\fbox{\xymatrix@=8pt{
\LTOO{String}\\
\LTO{N}\ar[u]^{\sf name} \ar@/^/[d]^{\sf age} \ar@/_/[d]_{\sf salary}\\
\DTOO{Int}}}}
\end{align*}
\end{centering}

\hrulefill
\vspace{.05in}

\begin{footnotesize}

\begin{tabular}{ccc}
\multicolumn{3}{c}{{\sf N1}} \vspace{.01in} \\
 {\sf ID}  &  {\sf name}  &  {\sf salary}  \\\hline 
1&Alice&\$100\\\hline 
2&Bob&\$250\\\hline 
3&Sue&\$300\\
\end{tabular}
\hspace{.25in}
\begin{tabular}{cc}
\multicolumn{2}{c}{{\sf N2}} \vspace{.01in} \\
 {\sf ID}  &   {\sf age}   \\\hline 
4&20\\\hline 
5&20\\\hline 
6&30\\
\end{tabular}
\hspace{.21in}
$\Fromm{ \llbracket \Delta_F \rrbracket }$
\hspace{.21in}
\begin{tabular}{cccc}
\multicolumn{4}{c}{{\sf N}} \vspace{.01in} \\
 {\sf ID}  &   {\sf name}   &   {\sf salary}   &   {\sf age}   \\\hline 
1&Alice&\$100&20\\\hline 
2&Bob&\$250&20\\\hline 
3&Sue&\$300&30\\
\end{tabular}

\vspace{.05in}
\hrulefill
\vspace{.05in}

\begin{tabular}{ccc}
\multicolumn{3}{c}{{\sf N1}} \vspace{.01in} \\
 {\sf ID}  &  {\sf name}  &  {\sf salary}  \\\hline 
1&Alice&\$100\\\hline 
2&Bob&\$250\\\hline 
3&Sue&\$300\\
\end{tabular}
\hspace{.25in}
\begin{tabular}{cc}
\multicolumn{2}{c}{{\sf N2}} \vspace{.01in} \\
 {\sf ID}  &   {\sf age}   \\\hline 
4&20\\\hline 
5&20\\\hline 
6&30\\
\end{tabular}
\hspace{.21in}
$\Too{ \llbracket \Sigma_F \rrbracket }$
\hspace{.21in}
\begin{tabular}{cccc}
\multicolumn{4}{c}{{\sf N}} \vspace{.01in} \\
 {\sf ID}  &   {\sf name}   &   {\sf salary}   &   {\sf age}   \\\hline 
1&Alice&\$100& {\sf age}(1)\\\hline 
2&Bob&\$250& {\sf age}(2) \\\hline 
3&Sue&\$300&  {\sf age}(3) \\\hline
4& {\sf name}(4) & {\sf salary}(4) & 20\\\hline 
5& {\sf name}(5) & {\sf salary}(5) & 20\\\hline 
6& {\sf name}(6) & {\sf salary}(6) & 30\\
\end{tabular}

\vspace{.05in}
\hrulefill
\vspace{.05in}

\begin{tabular}{ccc}
\multicolumn{3}{c}{{\sf N1}} \vspace{.01in} \\
 {\sf ID}  &  {\sf name}  &  {\sf salary}  \\\hline 
1&Alice&\$100\\\hline 
2&Bob&\$250\\\hline 
3&Sue&\$300\\
\end{tabular}
\hspace{.25in}
\begin{tabular}{cc}
\multicolumn{2}{c}{{\sf N2}} \vspace{.01in} \\
 {\sf ID}  &   {\sf age}   \\\hline 
4&20\\\hline 
5&20\\\hline 
6&30\\
\end{tabular}
\hspace{.21in}
$\Too{ \llbracket \Pi_F \rrbracket }$
\hspace{.21in}
\begin{tabular}{cccc}
\multicolumn{4}{c}{{\sf N}} \vspace{.01in} \\
 {\sf ID}  &   {\sf name}   &   {\sf salary}   &   {\sf age}   \\\hline 
1&Alice&\$100&20\\\hline 
2&Bob&\$250&20\\\hline 
3&Sue&\$300&20\\\hline
4&Alice&\$100&20\\\hline 
5&Bob&\$250&20\\\hline 
6&Sue&\$300&20\\\hline
7&Alice&\$100&30\\\hline 
8&Bob&\$250&30\\\hline 
9&Sue&\$300&30\\
\end{tabular}

\end{footnotesize}

\caption{Example Data Migrations (see Section~\ref{schmap})}
\label{fmd}
\end{figure}





%
%
%

\subsection{Mapping Composition}  

Given schema mappings $F: S \to T$ and $G: T \to U$, the {\it mapping composition} problem (Section~\ref{comp}) is to construct a schema mapping $G \circ F : S \to U$ that is equivalent with respect to query generation. 

In one sense, the mapping composition problem is trivial~\cite{patrick} for our formalism: $\Delta_{F \circ G} \cong \Delta_G \circ \Delta_F$, $\Pi_{F \circ G} \cong \Pi_F \circ \Pi_G$, and $\Sigma_{F \circ G} \cong \Sigma_F \circ \Sigma_G$.  But this solution is not wholly satisfactory because in practice a mixture of $\Delta,\Sigma, \Pi$ functors may be needed to accomplish any particular task (similarly, in SQL a mixture of joins and unions may be needed to accomplish any particular task).  The following results are proved in~\cite{patrick} and~\cite{relfound}:
\begin{itemize}
\item Every composition $\Sigma_F \circ \Delta_G$ is isomorphic to $\Delta_{F'} \circ \Sigma_{G'}$ for some $F',G'$.   This statement is also true if $\Sigma_F$ is replaced with $\Pi_F.$  \\

\item Pairs of the form $(F, G)$, denoting $\Sigma_F \circ \Delta_G$, are closed under composition.  This statement is also true if $\Sigma_F$ is replaced with $\Pi_F$.  Such pairs can be specified in an intuitive ``select-from-where'' syntax, described in~\cite{patrick} and~\cite{jfp}. \\

\item Triples of the form $(F, G, H)$, denoting $\Sigma_F \circ \Pi _G \circ \Delta_H$, are closed under composition, provided that $F$ is a {\it discrete op-fibration}~\cite{BW}, which is exactly the ``union compatibility'' condition~\cite{Doan:2012:PDI:2401764} that $\Sigma_F$ performs unions over tables whose columns match; Figure~\ref{fkmex} is not a discrete op-fibration. 
 \end{itemize}

\subsection{Mapping Inversion}  
Given a schema mapping $F : S \to T$, the {\it mapping inversion} problem (Section~\ref{secinv1}) is to construct a mapping $F^{-1} : T \to S$ that somehow ``undoes'' $F$.

Our formalism has strong inversion properties but does not have inverses per se.  When there exists $F^{-1} : T \to S$ such that $F \circ F^{-1} = id$ and $F^{-1} \circ F = id$, then $\Delta_F \circ \Delta_{F^{-1}} \cong id$,  $\Sigma_F \circ \Sigma_{F^{-1}} \cong id$, and  $\Pi_F \circ \Pi_{F^{-1}} \cong id$.  In general $F$ need not have an inverse, but when $S$ and $T$ have finite initial algebras / term models (which is a priori undecidable, and implies decidability of $S$ and $T$) it is possible to construct $F^{-1}$ whenever it exists by considering all possible functors $T \to S$.  When $F$ has a right adjoint $G : T \to S$, a weaker condition than having an inverse, there are canonical morphisms $\Sigma_F \to \Delta_G$ and $\Delta_F \to \Pi_G$.  

In practice ``round-tripping''~\cite{Doan:2012:PDI:2401764} of data is desirable even when inverses do not exist.  For example, projection, because it forgets information, typically cannot be inverted, but we may want to remember where the projected data originated.  In our formalism the adjunctions between $\Sigma$,$\Delta$,$\Pi$ provide round-tripping.  For example, for every  $F:S\to T$ and $S$-instance $I$ there is a canonical morphism $I \to \Delta_F(\Sigma_F(I))$, the {\it unit} of the $\Sigma_F \dashv \Delta_F$ adjunction, which describes where each ID in $I$ is sent to by $\Sigma_F$ (and similarly for $\Pi_F$).  Dually, for every $T$-instance $J$ there is a canonical morphism $\Sigma_F(\Delta_F(J)) \to J$, the {\it co-unit} of the $\Sigma_F \dashv \Delta_F$ adjunction, which describes where the IDs in $\Delta_F(J)$ originate (and similarly for $\Pi_F$).   The unit and co-unit can be used to obtain, for every morphism $h : \Sigma_F(I) \to J$, a mate $h' : I \to \Delta_F(J)$ and vice-versa (and similarly for $\Pi_F$).  Relating adjointness to existing relaxed notions of inverse such as quasi-inverse~\cite{Fagin:2008:QSM:1366102.1366108} is an important area for future work.

\subsection{Schema matching}  

Given database schemas $S$ and $T$, the {\it schema matching} problem (Section~\ref{matching}) is to automatically suggest schema mappings $S \to T$ to the user. 

 In this section, we define two schema matching techniques used by CQL.  Our techniques compare entities, and foreign keys and attributes (``symbols'') by name, as strings, and so our techniques depend on having (probably user-provided) names whose similarity as strings reflects their semantic similarity.  Let $\sigma : {\sf String},{\sf String} \to [0,1]$ be any string similarity function~\cite{Doan:2012:PDI:2401764} where a value of $1$ indicates a ``good'' match and a value of $0$ indicates a ``bad'' match.  


\begin{itemize}
\item  The first technique attempts to infer a schema mapping $F : S \to T$.  For each entity $s \in S$, we define $F(s) := t$ where $t \in T$ is an entity that maximizes $\sigma(s,t)$.  For each symbol $f : s \to s' \in S$, we then consider the set $X$ of symbols $F(s) \to F(s')$.  If $X$ is non-empty, we choose a symbol $g \in X$ that maximizes $\sigma(f,g)$ and set $F(f) := g$.  If $X$ is empty but there is a shortest path $p$ from $F(s)$ to $F(s')$, we set $F(f) := p$.  If no shortest path exists, the match fails.  The $F$ so constructed is only a candidate schema mapping: CQL must verify that $F$ preserves provable equality in $S$.  \\ 

\item The second technique attempts to infer a schema $A$ and schema mappings $F : A \to S$ and $G : A \to T$.  Such a span of mappings can be interpreted as a query of the form $\Sigma_F \circ \Delta_G$ or $\Pi_F \circ \Delta_G$. Let $c$ be some user-provided string similarity cutoff.  The entities of $A$ are those pairs of $S$-entities and $T$-entities $(s,t)$ such that $\sigma(s,t) > c$. The symbols $(s,t)\to(s',t')$ of $A$ are those pairs of $S$-symbols and $T$-symbols $(f : s \to s', g : t \to t')$ such that $\sigma(f,g) > c$.  The mappings $F$ and $G$ are projections. \\


\end{itemize}

\vspace{-.4in}
\section{Conclusion}
\label{concl}

When comparing our algebraic approach to model management with other approaches originating in relational database theory~\cite{DBLP:books/aw/AbiteboulHV95} it is important to note that our databases are ``deductive databases''~\cite{DBLP:books/aw/AbiteboulHV95}.  That is, we define databases ``intensionally'', as sets of equations, rather than as sets of tables.  As such, care must be taken when mediating between our definitions and relational definitions.  For example, our instances can be ``inconsistent'' in the sense that an instance can prove $1=2$ for two distinct constant symbols $1$ and $2$.  Such situations are often, but not always~\cite{Ghilardi06didi}, errors, and the CQL tool checks for such situations using standard techniques based on ``conservative theory extensions''~\cite{Ghilardi06didi}.  In addition, our schemas do not define a set of constants (a ``domain'') that all the instances on that schema share, as is customary in relational database theory~\cite{FKMP05}.  Hence our approach is closer in spirit to traditional logic~\cite{opac-b1104628} than database theory~\cite{DBLP:books/aw/AbiteboulHV95}.

There are many connections between our algebraic approach to model management and the ED-based approach described in Section~\ref{mm}.  EDs are more expressive than our purely equational data integrity constraints and can be added to our formalism in a simple way, described in~\cite{SPIVAK_2014} (although in~\cite{SPIVAK_2014}, EDs are called ``lifting problems'').  In ED-based approaches the ``chase''~\cite{Doan:2012:PDI:2401764} operation has a similar semantics to our $\Sigma$ operation, and a formal comparison between the chase and $\Sigma$ is forthcoming.  


{\bf Acknowledgements. } The authors thank Lucian Popa, Eswaran Subrahmanian, and Peter Gates and were supported by NIST SBIR grant 70NANB 16H178,  AFOSR grant FA9550--14--1--0031 and NASA grant NNL14AA05C. 

{\bf This paper appears in WADT 2016: Recent Trends in Algebraic Development Techniques, pp 56–69.}


 \bibliographystyle{splncs03}
 \bibliography{newbib}
\end{document}